\documentclass[aps,prl,twocolumn]{revtex4}
\usepackage[pdftex]{graphicx}
\usepackage{amssymb,amsfonts,amsmath}
\usepackage{color}

\begin{document}

\author{Slava V. Rotkin$^{1*}$, Vasili Perebeinos$^2$, Alexey G. Petrov$^3$, and Phaedon Avouris$^2$}

\affiliation{ $^1$ Physics Department, Lehigh University, 16
Memorial Dr.~E., Bethlehem, PA 18015; \\
 Center for Advanced Materials and Nanotechnology,
 Lehigh University,
 5 E.~Packer Ave., Bethlehem, PA 18015. E-mail: rotkin@lehigh.edu\\
$^2$IBM Research Division, T. J. Watson Research Center,
Yorktown Heights, New York 10598 \\
$^3$Ioffe Institute, 26 Polytekhnicheskaya, St. Petersburg,
194021, Russia}


\title{An essential mechanism of heat dissipation in carbon nanotube electronics}
\date{\today}



\begin{abstract}
Excess heat generated in integrated circuits is one of the major
problems of modern electronics. Surface phonon-polariton
scattering is shown here to be the dominant mechanism for  hot
charge carrier energy dissipation in a nanotube device fabricated
on a polar substrate, such as $SiO_2$.  Using microscopic quantum
models the Joule losses were calculated for the various energy
dissipation channels as a function of the electric field, doping,
and temperature. The polariton mechanism must be taken into
account to obtain an accurate estimate of the effective thermal
coupling of the non-suspended nanotube to the substrate, which was
found to be 0.1-0.2 W/m.K even in the absence of the bare phononic
thermal coupling.
\end{abstract}


\keywords{heat dissipation,thermal conduction,nanotube
electronics,hot electron relaxation,energy relaxation rate,surface
phonon-polariton}


\maketitle

\section*{Introduction}

Nowadays the vast majority of human activities are supported by
advanced information technology which cannot become a
transformative power without progress in semiconductor
electronics. Exponentially increasing  dissipated power density in
integrated electronics circuits is one of the ``grand challenges"
of modern electronics \cite{heat-in-ic,crchandbook}. Excess heat
generated by electric currents in the elements of circuits cannot
be completely removed because of the high power density, which
leads elevated device operation temperatures, performance
reduction, and ultimately results in hardware failures.

Heat dissipation takes place via the transfer of thermal energy that
is generated by hot charge carriers, to the cooling units
eventually. The three classical mechanisms of heat transfer are
convection, conduction, and radiation. The former is inapplicable in
existing circuit architecture having no fluid or gas flowing inside.
The latter has a total emissive power proportional to $T^4$
according to the Stefan–-Boltzmann law, and is not very effective
within the acceptable range of temperatures
\cite{SBnearfield,fnote1}. Thus the major mechanism responsible for
the device cooling is the thermal conductance. High thermal
conductance can be achieved in bulk solids, but even a small vacuum
gap is a quite effective thermal insulator.

This is the case for transistor channels involving a
nanostructure, for example a nanowire or a
nanotube\cite{avouris-dev,lieber1,lieber2}, which is not
chemically connected to the substrate, instead it is bound by
relatively weaker van der Waals forces. This weak binding has a
positive effect on the mobility of the charge carriers, but also
has a negative effect on the thermal conductance. In existing
nanotube transistors only the contacts have a good thermal
exchange rate while the thermal conductance over the van der Waals
separation gap is very low, of the order of 0.05--0.2 $W/K.m$
\cite{therm-cond-exp-theor1,therm-cond-exp-theor2,therm-cond-exp-theor3,therm-cond-exp-theor4,therm-cond-exp-theor5}.
In addition, the total thermal conductance of the contacts is not
too high, being proportional to the cross-sectional area of the
nanotube.

In this work we propose that a fourth mechanism of heat
dissipation from the hot charge carriers in the nanotube channel
into a polar $SiO_2$ substrate should operate and argue that such
a mechanism would dominate the thermal transfer especially at
large device currents (high bias voltages). Understanding the
importance of such a mechanism invokes concepts that are specific
to carbon nanotube (NT) field-effect transistors (FET): low
thermal conductance due to the van der Waals separation gap (and
possible surfactant coating) and the small contact area, high
Fermi velocity of the NT charge carriers and the existence of the
special electromagnetic (EM) surface modes in the vicinity of the
NT channel. Hot electrons, as we will discuss below, can excite
surface polariton modes. This mechanism while important for
nanoscale channels is ineffective in a macroscopic system because
the EM field of the polariton decays exponentially with the
distance from the surface plane \cite{polder}. This leads to a
power low scaling of the scattering rate with the distance from
the surface plane \cite{p-subm}. A surface polariton mode would
quickly dissipate into bulk substrate modes, thus providing a heat
transfer mechanism with an efficiency exceeding the standard
thermal conductance or radiation mechanisms.


The existence of surface polariton modes was predicted in 1899
\cite{sommerfeld} and were later observed at AM radio-wave
frequencies. In condensed matter, a metal surface is known to
support surface EM modes that is plasmon-polaritons; the focus of
recent plasmonics studies \cite{plasmonics}. The plasmon-polariton
EM mode exists if the real part of the dielectric function
$\varepsilon$ of the substrate equals negative one \cite{mahan}.
Similarly, one can observe a surface polariton in a polar insulator
where a strong optical phonon mode provides the required condition:
$\varepsilon=-1$. Then the frequency of the surface phonon-polariton
(SPP) is close to the frequency of the optical phonon. For example,
for a SiO$_2$ surface the optical phonon modes that can support SPP
are at $\hbar\omega^{(\nu)}=$
 50, 62, 100 and 149 meV
\cite{quartz1,lynch,spitzer}.~
Hot charge carriers in a NT FET with energy exceeding the
polariton energy would effectively release energy into the SPP
channel.

We stress that despite the fact that this mechanism is similar to
a classical one for hot charge carrier energy relaxation whereby
the energy dissipates into optical phonons of the NT
\cite{kane,javey,park,perebeinos}, the SPP mechanism transfers the
thermal energy directly into a bulk substrate, while the classical
one only redistributes it between the NT electrons and the NT
lattice, to be followed by thermal conduction into the leads
and/or the substrate. The SPP mechanism also differs from the
classical radiation since the charge carriers in the channel
couple to the evanescent EM modes only. The greatly enhanced
surface electric field of such modes has much larger overlap with
the NT channel than it would be in the case of the free vacuum EM
modes\cite{fnote1}.

\section*{Simulation approach}

\begin{figure}[tb]
\centering
 {   \includegraphics[width=3in]{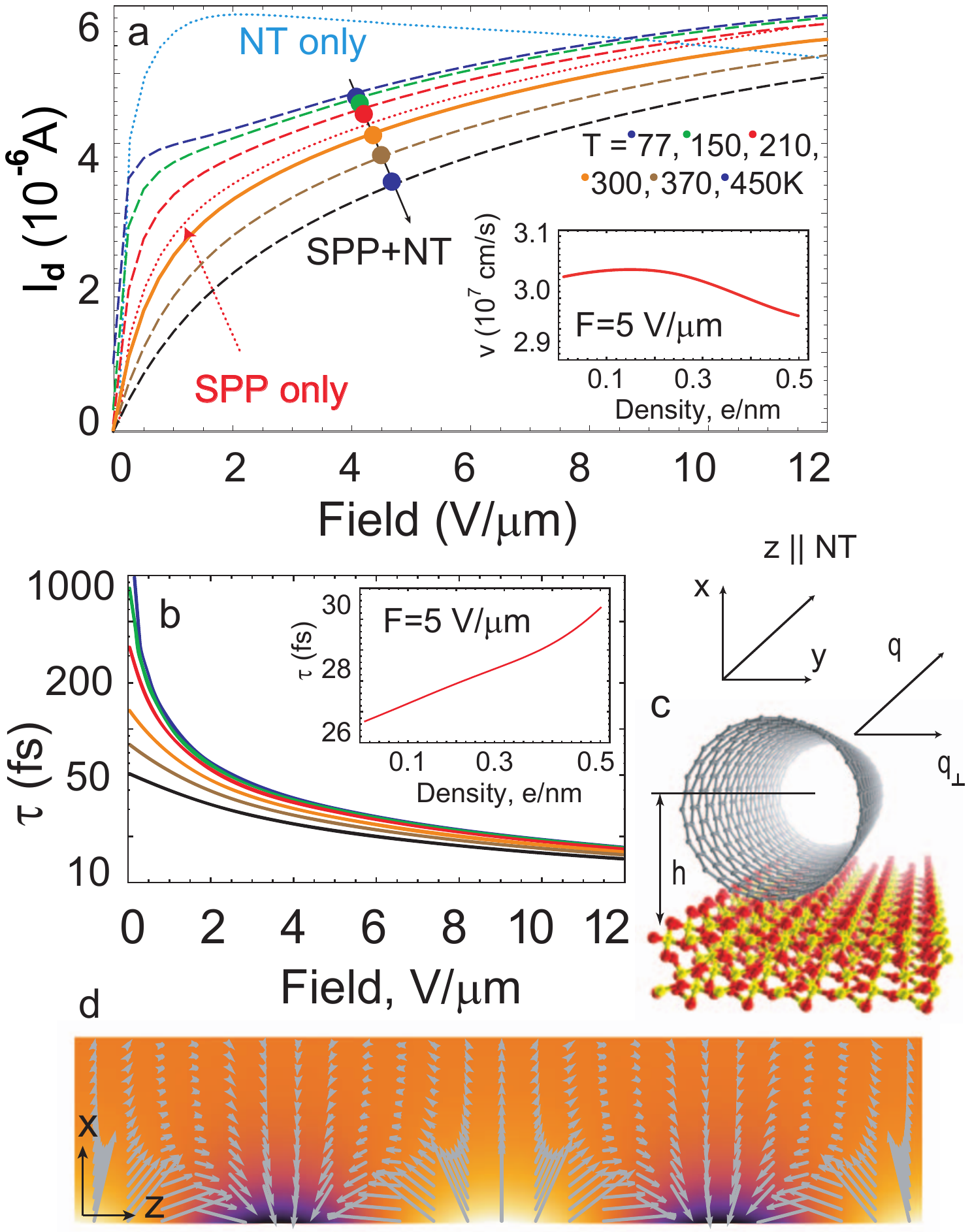}
  \caption{Role of SPP inelastic
scattering mechanism for NT FET transport.
 (a) Current-electric field curves for [17,0] NT at the doping level
$\rho=0.1~e/nm$, $T= 77, 150, 210, 300, 370, 450~K$ as indicated
by the color code. Full line corresponds to the room temperature
characteristic with both SPP and NT phonon scattering mechanisms
included, as compared to the SPP scattering only (red dotted curve
slightly above) and NT phonon scattering only (light blue dotted
curve at the top). Insets in (a) shows the drift velocity vs. the
doping level at $F=5~V/\mu m$ and $T=300$~K.
 (b) The energy
relaxation time vs. applied electric field using both SPP and NT
scattering for the same NT and same temperatures. Inset in (b)
shows dependence of the energy relaxation time on the doping level
at $F=5~V/\mu m$ and $T=300$~K.
 (c) Schematics of the NT channel on a SiO$_2$ substrate (3D image is
generated using Molecular Dynamics in NAMD/VMD). Accepted
convention for the components of the in-plane wave-vector of the
SPP mode and the coordinate system are shown.
 (d) Distribution of the electric potential and its gradient for a
Surface Phonon Polariton (SPP) mode. }
  \label{fig-one-two}
  }
\end{figure}

We have adopted the model described in detail in the previous
works \cite{perebeinos,p-subm,perebeinos2}. In brief, a standard
tight-binding Hamiltonian for the NT charge carriers, $H_e=\sum
t_{ij} a^\dagger_i a_j$, is supplemented with Su-Schrieffer-Heeger
terms for the electron-NT-lattice-phonon coupling (Eq.(1) of
Ref.\cite{perebeinos}):
\begin{equation}
H_{NT}=\sum \limits_{k,q,\mu} M^\mu_{kq}
\left(a^\dagger_{-q\mu}+a_{q\mu}\right)
\left(c^\dagger_{v,k+q}c_{v,k}-c^\dagger_{c,k+q}c_{c,k}\right)
 \label{ssh}
\end{equation}
here the matrix element of the interaction is $M^\mu_{kq}\propto
\Xi=5.3$~eV/\AA. The coupling constant is $\Xi\propto t-t_o$,
where $t_o=3$~eV is the bare tight-binding hopping integral being
modulated by the phonon modes. $a_{q\mu}$ and $c_{c/v,k}$ are
annihilation operators for NT phonons ($\mu$ is a phonon branch
index) and charge carriers in conduction and valence bands,
respectively, that are labeled with the one-dimensional axial
momentum $q$ or $k$. In addition to the NT lattice phonons we
include the operator of the interaction with the SPP mode (cf.
Eq.(2) in Ref.\cite{jetpl}):
\begin{equation}
H_{SPP}=\sum \limits_{k,q,\nu} V^\nu_{k,q,q_{\perp}}
\left(a^\dagger_{-q,q_{\perp},\nu}+a_{q,q_{\perp},\nu}\right)
c^\dagger_{k+q}c_{k}
 \label{spp}
\end{equation}
where $a_{q,q_{\perp},\nu}$ is the annihilation operator for the
$\nu$-th surface phonon mode (not to be confused with the NT
lattice phonons),
and an explicit form of 
$V^\nu_{k,q,q_{\perp}}$ was derived in Ref.\cite{jetpl}:
\begin{equation}
V^\nu_{k,q,q_{\perp}}= \frac{2i\pi \, e{\cal F}_{\nu} \,(-q)^{m}
    e^{-h\sqrt{q^2+q_{\perp}^2}}I_{m}(|qR|)}{\sqrt[4]{q^2+q_{\perp}^2}
    \left(\sqrt{q^2+q_{\perp}^2}-q_{\perp}\right)^{ m}}
 \label{Vspp}
\end{equation}
here $q$ and  $q_{\perp}$ are the components of the SPP momentum
that are parallel and normal to the NT axis respectively (see
Fig.\ref{fig-one-two}c), $m$ is the angular momentum transfer to
the SPP mode (equals 0 in case of the NT intra-subband scattering
which is the strongest scattering channel), $R$ is the NT radius,
$h\simeq R+4$~\AA~ is the distance from the NT axis to the
surface, $I_m(x)$ is the Bessel function of imaginary argument
\cite{abramovitz}. $e$ is the elementary charge and ${\cal
F}_{\nu}$ characterizes the strength the electric field of the SPP
mode \cite{mahan}:
\begin{equation}
    {\cal F}^2_{\nu}=\frac{\hbar\omega_{LO}^{(\nu)}}{2\pi S}
\sqrt{\frac{1+\epsilon_0^{-1}}{1+\epsilon_\infty^{-1}}}\,
    \left(\frac1{\epsilon_\infty+1}-\frac1{\epsilon_0+1}\right)
 \label{F-factor}
\end{equation}
where $S$ is the normalization surface area, $\omega_{LO}^{(\nu)}$
is the frequency of the longitudinal optical phonon of the polar
substrate with the static permittivity $\epsilon_0$ and the
low-frequency dielectric function $\epsilon_\infty$. We should
emphasize that while there are four surface modes present (the
highest energy one being doubly degenerate), only two of them have
strong coupling strength \cite{p-subm,jetpl}: $S{\cal
F}^2_{\nu=1,..5}=0.042, 0.38, 0.069, 1.08, 1.08$. The
characteristic distribution of the electric field of the SPP mode
above the substrate surface is shown in Fig.\ref{fig-one-two}d.
The field oscillates along the surface (along $\vec q$) and decays
exponentially in the space (along $\vec q_\perp$). The transverse
component of the SPP momentum, $q_\perp$, is not conserved and has
to be integrated for the final result.

To simulate the current-voltage curves we solved numerically the
steady-state Boltzmann equation for the NT in a constant electric
field (including as many subbands as needed for given temperature
and drain field values). One can assume that in the bulk of the
FET channel the drain voltage $V_d$ is related to the applied
electric field via the effective channel length $L$: $F=V_d/L$.
The scattering rates for the Boltzmann equation are as follows
(see Refs.\cite{p-subm} and \cite{jetpl} for further details):
\begin{eqnarray}
    W_{kq}=&\displaystyle\frac{S}{\pi\hbar}\frac{2\pi}{L}\sum\limits_{\nu}
     \int_0^\infty dq_\perp |V^\nu_{k,q,q_{\perp}}|^2\times
   \nonumber\\
&\times\left(n_{q\nu}\,
\delta\left(E_f(k)+\hbar\omega_{SO}^{(\nu)}-E_i(k+q)\right)+
\right.
   \nonumber\\
 +&\left.(n_{-q\nu}+1)\delta\left(E_f(k)-E_i(k+q)-\hbar\omega_{SO}^{(\nu)}\right)\right)
    \label{eqn:golden}
\end{eqnarray}
here $\hbar\omega_{SO}^{(\nu)}$ is the frequency of the SPP mode
\cite{mahan}; $n_{q\nu}$ its phonon occupation number,
$E_{f/i}(k)$ is the energy of the final/initial state as obtained
from the Hamiltonian $H_e$. Non-equilibrium distribution function
$g_k$ has been calculated numerically for every given value of the
temperature, field and charge density and then it was used to
compute the total electron current, $I_d$, as well as to determine
the partial phonon emission rates for the heat dissipated in the
NT lattice $P_J$ and that dissipated directly into the substrate
via the SPP mode $P-P_J$, according to:
\begin{equation}
P-P_J=\frac{2}{\hbar L}\sum_{k,q}W_{k,q}
g_k\left(1-g_{k+q}\right)\left(E(k)- E(k+q)\right)
\label{eqn:losses}
\end{equation}
where  $P=I_dF$ are the total Joule losses. A similar equation
holds for NT losses $P_{J}$ with the scattering rate obtained from
Eq.~(\ref{ssh}). In the following we study dependencies of the SPP
to the NT channel loss ratio, defined as $\xi=P/P_J-1$, on the
bias, density, and temperature.


\section*{SPP Scattering and NT-FET High Bias Regime}

The role of the SPP 
modes with respect to charge carrier scattering (so-called Remote
Interface Phonon scattering) for Si devices has been first studied
by Hess and Vogl \cite{vogl}, though this mechanism was not found
to dominate the transport \cite{siRIP}. As we show in this work
, for nanotube devices the SPP scattering mechanism is very fast,
as been indicated already in our earlier paper \cite{jetpl} and
cannot be neglected even for the low-field transport
\cite{p-subm}.

  The SPP energy relaxation rate exceeds the
intrinsic NT phonon relaxation rates (for both optical and
acoustic modes) for the whole range of the charge densities and
applied electric fields (gate and drain biases) studied here as
shown by the results in Fig.~\ref{fig-one-two}b.

Recently SPP scattering was demonstrated to determine the mobility
in graphene on a polar substrate and to limit the ultimate
performance of graphene devices
\cite{fuhrer-n-theory1,fuhrer-n-theory2,ken-n}.
Fig.\ref{fig-one-two}a shows typical current-electric field curves
calculated with and without the SPP mechanism taken into account.
Neglecting the SPP channel significantly overestimates the
low-field current. Two important conclusions are drawn from these
data: firstly, the negative differential resistance region of the
current-voltage characteristic is totally removed when the SPP
channel included, because the very effective SPP energy relaxation
prevents the hot electron run-away as we further discuss in Ref.
\cite{p-subm}. Secondly, the SPP mechanism dominates over the NT
phonon scattering for both low- and high-field regimes but the
remnants of the non-SPP behavior can be seen at low applied field
especially at low temperature.

Results on the relative importance of the SPP and NT-phonon
scattering channels for the energy relaxation of the hot charge
carriers are shown in Fig.\ref{fig-three}. In
Fig.\ref{fig-three}(a) we plot, in a logarithmic scale, the
partial specific energy losses due to the SPP channel (full
curves) and only the NT phonon channel (broken curves), calculated
for [17,0] NT at the doping level $\rho=0.1~e/nm$ at various
temperatures. The SPP losses are up to two orders of magnitude
larger than the NT phonon losses. Fig.\ref{fig-three}(b) presents
the ratio of the SPP partial losses to the NT phonon partial
losses calculated for the same NT. The high frequency NT optical
phonons increase the NT partial losses at higher fields and higher
$T$ (Fig.\ref{fig-three}b). Even in that limit their ratio $\xi$
is always greater than one.

We also studied the dependence of the loss ratio on the doping
level $\rho$ (Fig.\ref{fig-three}e, inset) at the fixed applied
electric field $F=5~V/\mu m$. The relative decrease of the SPP
contribution at higher electron density (higher gate bias) is
shown in Fig.\ref{fig-three}(e) as a percentage of the total
losses and is due to the SPP and NT phonon losses having a
different functional dependence on $\rho$. The non-equilibrium
distribution function (obtained as a numerical solution of the
Boltzmann equation) varies with the doping level. The SPP losses,
correlated with the non-equilibrium distribution function, scale
similar to the total losses (proportional to the current) and are
approximately a linear function of the doping level (total number
of electrons). The NT phonon losses are superlinear in $\rho$ and
grow faster, thus their ratio $\xi$ decreases with $\rho$.

To understand the dependencies of the NT and SPP losses on field
strength and temperature, we show the scattering rate of electrons
via these two channels in different bands as a function of energy
(see Supplemental Information Fig.~3). In the first two energy
bands the SPP scattering dominates over the NT scattering, while
in the higher energy bands the NT scattering becomes comparable to
that of the SPP scattering. This is because in the SPP scattering
the angular momentum is conserved $\Delta m=0$, so that only
intra-band scattering takes place, while NT optical phonons can
lead to both intra and inter-band scattering and the phase space
for such scattering grows with the band index. Therefore, as the
electronic distribution becomes hotter either due to the field or
temperature the relative role of the NT phonons increases and the
loss ratio $\xi$ decreases. 
Fig.~4S (Supplemental Information) further supports that the
electronic distribution function rather than the scattering rate
(see Eq.~(\ref{eqn:losses})) is primarily responsible for the
value of the loss ratio at the high bias regime. Therefore the NT
and SPP losses scale in exactly the opposite way with temperature.

\section*{Near-Field Thermal Conductance and Self-Heating of NT-FET Channel}



\begin{figure}[tb]
\centering
 {   \includegraphics[width=3in]{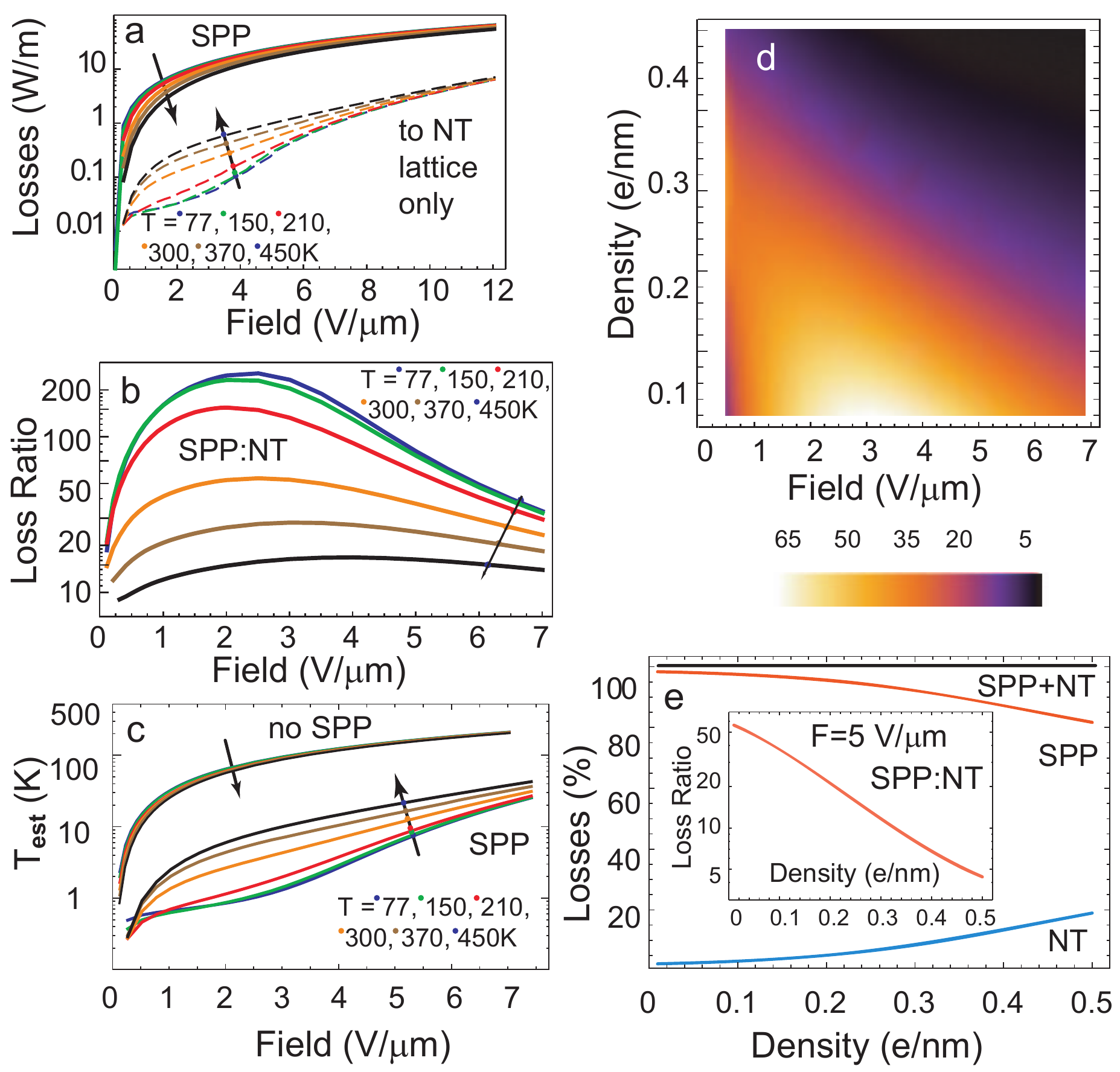}
  \caption{Comparison of the hot electron energy relaxation via SPP and other
inelastic scattering channels.
 (a) Logarithm of the partial
specific energy losses due to the SPP channel (full lines) and the
NT phonon channel (broken curves) for [17,0] NT at the doping
level $\rho=0.1~e/nm$ (the temperature range is indicated by the
color code $T= 77, 150, 210, 300, 370, 450$~K arrows point to
increasing temperatures).
 (b) The ratio of the SPP
partial losses to the NT partial losses, $\xi=P/P_J-1$, vs.
applied electric field for the same NT.
 (c) Estimated effective NT FET channel temperature vs. applied electric field
calculated for [17,0] NT on SiO$_2$ substrate. NT phonon
contribution (upper curves' family) is compared with both NT and
SPP channels (lower curves' family) for the bare thermal coupling
to the substrate $g_o=0.19~\frac{W}{K.m}$, $\rho=0.1~e/nm$.
 (d) The loss ratio $\xi$, calculated including self-consistent NT heating, is
shown as a function of the doping level and the applied electric
field. Here $g_o=0.19~\frac{W}{K.m}$, $T_{sub}=300~K$.
 (e) Relative contributions of the SPP (red curve) and
NT (blue curve) channels to the total energy losses (black curve)
vs. the doping level at $F=5~V/\mu m$; the inset shows their
ratio.}
  \label{fig-three}
  }
\end{figure}


\begin{widetext}

\begin{figure}[tb]
\centering
 {   \includegraphics[width=6in]{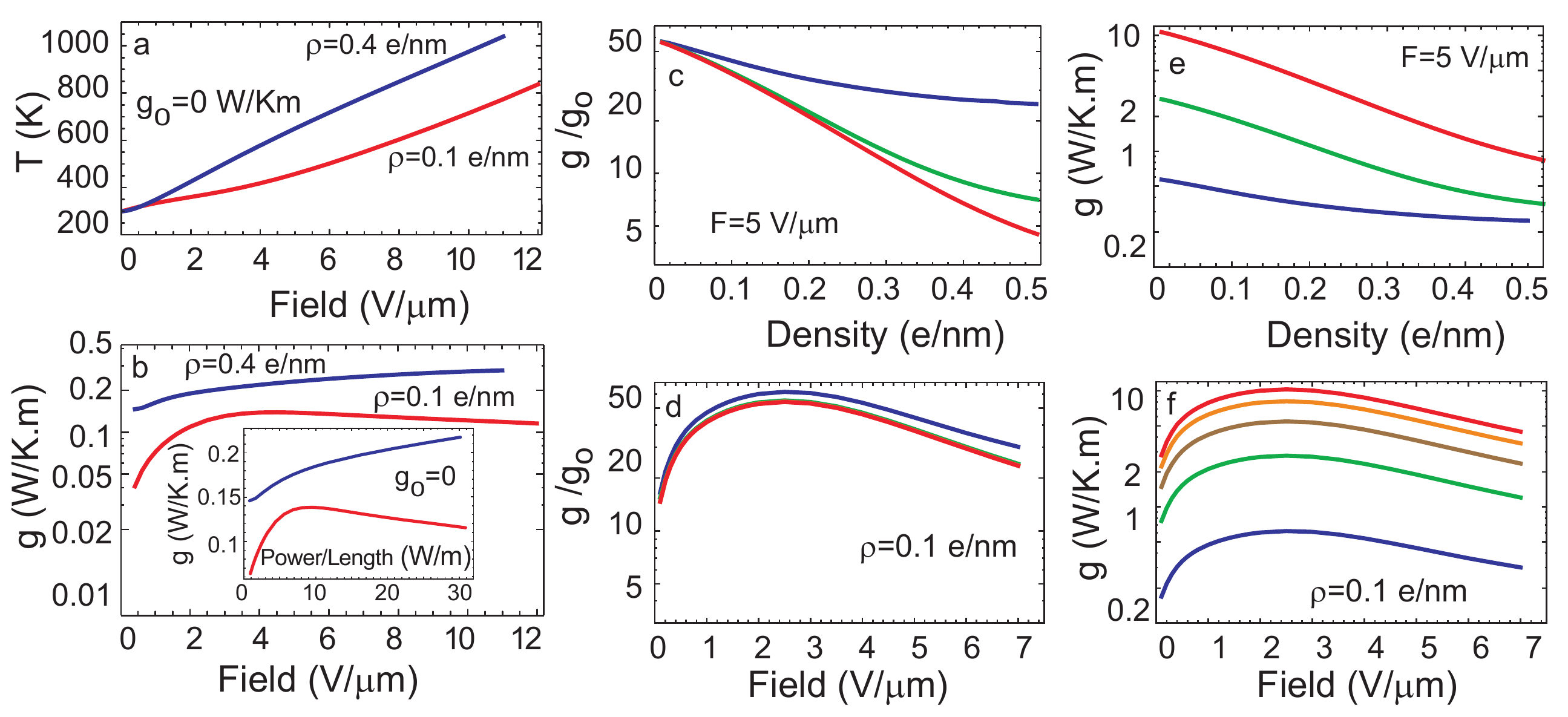}
  \caption{SPP as a heat dissipation mechanism.
  (a) Self-consistent effective NT FET channel
temperature vs. applied electric field calculated for [17,0] NT on
SiO$_2$ substrate, $T_{sub}=300~K$, the doping level is
$\rho=0.4~e/nm$ (purple, upper) and $0.1~e/nm$ (red, lower). A
bare thermal coupling is set to $g_o=0$.
    (b) Logarithm of the effective thermal conductivity via the NT-SiO$_2$ interface for
the same NT as in (a) due to the SPP coupling, as extracted from
the self-consistent temperatures data, vs. applied electric field.
Inset shows the same data vs. dissipated power per NT length.
    (c-f) Effective thermal coupling $g$ and $g/g_o$-ratio
vs. the doping level at $F=5~V/\mu m$, (c) and (e); the same vs.
applied electric field at $\rho=0.1~e/nm$, (d) and (f). Bare
thermal coupling is $g_o=0.01, 0.05, 0.19~\frac{W}{K.m}$ in (c)
and (d) and $g_o=0.01, 0.05, 0.10, 0.15$ and $0.19~\frac{W}{K.m}$
(from blue to red) in (d) and (f).}
  \label{fig-foura}
  }
\end{figure}

\end{widetext}

Next we discuss another aspect of SPP scattering, namely we address
how the heat generation is controlled by the fast SPP energy
relaxation rates.

So far we set the NT temperature equal the substrate temperature.
That is we assumed an ideal thermal coupling to the substrate
which is a condition not confirmed experimentally. As a next level
of approximation we use thermal coupling values (per NT length)
from the literature
\cite{therm-cond-exp-theor1,therm-cond-exp-theor2,therm-cond-exp-theor3,therm-cond-exp-theor4,therm-cond-exp-theor5}
ranging from $g_o=0.05$ to $0.2~W/K.m$ and estimate the FET
channel heating. The lower bound of the thermal coupling, as it
will be shown below, gives so small a thermal exchange to the
substrate that it is almost equivalent to assuming no thermal
coupling at all (see Supplementary Figure 1). Thus we analyze
first the temperature of the NT FET channel for the high value of
the thermal coupling to the substrate $g_o=0.19~W/(K.m)$.

All energy loss of the hot electrons, same as the total dissipated
power, is distributed between two channels: the SPP  losses which
are to be subtracted from the total heat flux in the NT, and the
NT losses which have to be partially transmitted to the substrate
via the coupling $g_o$. Fig.\ref{fig-three}(c) presents an
effective channel temperature (the Joule overheating) as a
function of the applied electric field calculated for NT phonons
only (upper curves' family) and for both NT and SPP channels
(lower curves' family). The SPP scattering channel couples to the
substrate phonons and thus transfers most of the excess thermal
energy of the hot electron directly into the substrate. In
contrast, the NT phonon scattering transfers the energy to the NT
lattice. As a result the steady-state temperature of the NT
lattice increases from the ambient temperature, $T_{sub}$. For a
very long NT FET channel all edge effects due to the electron and
phonon thermal conductivity can be neglected, as well as a heat
flux due to the hot electron current into the drain electrode.
Given the thermal coupling rate we estimate the steady-state
temperature as:
\begin{equation}
T=T_{sub}+\Delta T=T_{sub}+ \frac{P_J}{g_o}
 \label{deltaT}
\end{equation}
where $P_J$ is the dissipated power per NT length due to the NT
phonon scattering. The Joule overheating $\Delta T$ is found to be
orders of magnitude smaller compared to the case when SPP channel
is neglected and all the measured Joule losses are attributed to
the NT phonons, i. e. $P_J=P=I_dF$ (see Fig.\ref{fig-three}c). The
ratio of the SPP losses to NT losses, $\xi$, as a function of the
applied field and doping level is given in Fig.\ref{fig-three}b
and in the inset (e). The temperature rise $\Delta T$ is inversely
proportional to $g_o$ according to Eq.(\ref{deltaT}). Even though
the loss ratio depends on $g_o$ (see below), which is not
precisely known, it is safe to conclude that the temperature of
non-suspended NT channel will be much lower for the same total
power density $P$.

Within this non-self-consistent scheme one overestimates the
overheating because one neglects $\Delta T$ when calculating the
current, the SPP and Joule heat (NT) losses. Next we part this
approximation and provide fully self-consistent analysis of the
channel temperature.

Given the expected temperature rise of the NT FET channel (for a
given value of the bare thermal coupling) we recalculate the
electron and phonon distribution and obtain the self-consistent NT
temperature by iteration. At $n$-th step we solve the Boltzmann
equation using the NT-lattice temperature $T_{n-1}$, calculated at
the previous step according to Eq.~(\ref{deltaT}), and then
compute the total energy released in the NT-phonon subsystem.
Next, using Eq.(\ref{deltaT}) we calculate the new lattice
temperature $T_{n}$. The process is iterated until convergence.

At zero bare thermal coupling $g_0=0$, SPP channel controls the
Joule heating. The field and concentration dependence of the FET
temperature at $g_o=0$ is presented in Fig.\ref{fig-foura}a. The
upper (purple) curve corresponds to [17,0] NT on SiO$_2$ substrate
at $T_{sub}=300~K$, $\rho=0.4~e/nm$. A temperature rise higher
than $600~K$ can be expected at such conditions \cite{submitted}.
By decreasing the doping level 4 times (red curve) we decrease the
steady-state temperature due to a smaller value of the Joule
losses, proportional to the current $\sim I_d\sim \rho$.
Additional decrease is associated with a gradual dependence of the
loss ratio $\xi$ on the density (Fig.\ref{fig-three}e inset). To
further explore the nature of this cooling down of the NT FET
channel we compute the effective thermal coupling to the SiO$_2$
substrate, defined as $g=(I_d F)/\Delta T$. We extract it from the
self-consistent temperatures data (as in Fig.\ref{fig-foura}a) and
plot it vs. the applied field, $F$, and vs. the total dissipated
power per NT length, $P$, in Fig.\ref{fig-foura}b.

In a real device the non-zero bare thermal coupling $g_o>0$
results in even higher values of the total thermal coupling.
Effective thermal coupling $g$ and the ratio $g/g_o$ are presented
in Fig.\ref{fig-foura}, panels (c-f) vs. the doping level at fixed
electric field $F=5~V/\mu m$ and also vs. the field for
$\rho=0.1~e/nm$ and various bare couplings $g_o=0.05, 0.10, 0.15$
and $0.19 \frac{W}{K.m}$ (from red to blue). We conclude that the
SPP, near-field, thermal conductance is always larger than the
bare thermal coupling by at least an order of magnitude.

We note that in the presence of dissipation mechanisms to both SPP
and NT channels, the total thermal conductivity of the NT FET
becomes a function of bias and density and the analysis of the
electric heating through Joule's, Fourier's and Ohm's laws, often
used in literature \cite{kuroda,pop}, must account for these
dependencies. In such a case, one has to calculate a
self-consistent non-equilibrium distribution function first and
find the lattice (phonon) temperature. The total thermal flux has
two components: the SPP channel flux and the NT-substrate flux. We
predict that for supported (non-suspended) NT device on a polar
substrate the SPP dissipation channel is major and dominates over
the NT-substrate thermal dissipation. Therefore, an effective
thermal coupling through the NT-substrate interface appears to be
significantly larger as compared to the bare thermal coupling on a
non-polar substrate.

\section*{Conclusions}

In this paper we presented a microscopic quantum modeling of a
novel heat dissipation mechanism for nanotube electronic devices.
This mechanism is specific for NT devices fabricated on polar
substrates, such as SiO$_2$, due to (i) existence of surface EM
modes at the frequencies of the surface phonon-polaritons, (ii)
strong coupling of such modes to the charge carriers in the NT
lying on the substrate. We note that a similar SPP thermal
coupling should exist in other 1D and 2D systems fabricated with
other channel materials on polar insulator substrates. In
particular, high-k oxides are expected to produce strong SPP
scattering due to their large Fr${\rm \ddot{o}}$hlich constants
and their  low optical phonon frequencies.

Using a semiempirical quantum approach we have calculated
current-voltage curves with and without SPP scattering and
concluded that the SPP mechanism dominates the scattering and
determines the drain current in the whole range of drain voltages,
and for all studied doping levels and temperatures. The current
(charge density) and temperature scaling of the SPP and NT
scattering mechanisms are different, thus allowing verification of
our predictions experimentally. In this work we focused on the FET
high bias regime and demonstrated that the SPP mechanism, being
much more efficient than the scattering by NT lattice modes,
results in the elimination of the negative differential resistance
part of the current-voltage curve.
~We analyzed the relative importance of the SPP channel, comparing
to all the other NT phonon mode scattering channels. In the whole
studied range the SPP channel dominates. Thus, most of the energy
losses are dissipated directly into the polar substrate and do not
contribute to the FET temperature rise (Joule overheating). We
showed that the SPP thermal coupling increases the effective
thermal conductance over the interface between the NT and such
polar insulator as $SiO_2$ by an order of magnitude. The
dependence of the effective thermal conductance of the NT/SiO$_2$
interface on the channel doping level was computed and may be used
to verify the model predictions. We note that plasmon-polaritons,
existing at a metallic surface, may also result in a strong
thermal coupling to the nanotubes.

\begin{acknowledgments}
SVR acknowledges partial support by DoD-ARL grant W911NF-07-2-0064
under Lehigh-Army Research Laboratory Cooperative Agreement, and
the Donors of the American Chemical Society Petroleum Research
Fund (ACS PRF 46870-G10).
\end{acknowledgments}


\end{document}